# Role of Vanadyl Oxygen in Understanding Metallic Behavior of $V_2O_5$ (001) Nanorods

*Raktima Basu,[*] Arun K. Prasad, Sandip Dhara,[*] and A. Das*

Nanomaterials and Sensors Section, Surface and Nanoscience Division, Indira Gandhi Centre for Atomic Research, Homi Bhabha National Institute, Kalpakkam−603 102, India

**Corresponding Authors**

[*]Email Address: raktimabasu14@gmail.com; dhara@igcar.gov.in

Telephone: +91 44 27480015

*Abstract*

Vanadium pentoxide ($V_2O_5$), the most stable member of vanadium oxide family, exhibits interesting semiconductor to metal transition in the temperature range of 530-560K. The metallic behavior originates because of the reduction of $V_2O_5$ through oxygen vacancies. In the present report, $V_2O_5$ nanorods in the orthorhombic phase with crystal orientation of (001), are grown using vapor transport process. Among three non-equivalent oxygen atoms in a $VO_5$ pyramidal formula unit in $V_2O_5$ structure, the role of terminal vanadyl oxygen ($O_I$) in the formation of metallic phase above the transition temperature is established from the temperature dependent Raman spectroscopic studies. The origin of the metallic behavior of $V_2O_5$ is also understood due to the breakdown of $pd\pi$ bond between $O_I$ and nearest V atom instigated by the formation of vanadyl $O_I$ vacancy, confirmed from the downward shift of the bottom most split-off conduction bands in the material with increasing temperature.

## 1. Introduction

One dimensional transition metal oxides exhibit unique structure-property relationships which help in developing new electronic and photonic devices.[1,2] Vanadium is a transition metal ([Ar]$3d^34s^2$) with multiple oxidation states leading to various stoichiometric oxides. Vanadium pentoxide ($V_2O_5$), which is essentially a semiconductor at room temperature, is the most stable form among them. Although, most of vanadium oxides such as $VO_2$, $V_2O_3$, $V_6O_{13}$ exhibit metal to insulator transitions (MITs) as a function of temperature, in case of $V_2O_5$ it is quite contentious to use the term 'MIT'.[3] However, there are few reports about transition to metallic phase of $V_2O_5$ films around 530 to 553K and surface metallicity of (001) facet around 340 to 400K.[4-6] The MIT is also reported to be reversible.[6] The metallic behavior makes $V_2O_5$ applicable as gas sensors,[7,8] thermochromic devices,[9,10] optical and electrical switches[11] around the transition temperature.

The origin of the metallic behavior, however, is still not fully understood. The metallic transition was reported because of the reduction of $V_2O_5$ to other lower ordered stoichiometric or non-stoichiometric oxides without any structural change.[5,12] In the structure of $V_2O_5$ there are three differently coordinated O atoms, namely, $O_I$ (vanadyl), $O_{II}$ (bridging), and $O_{III}$ (chain) (schematic in the supplementary information Figure S1).[13] There is also a dispute in the identification of the oxygen responsible for the reduction as all three $O_I$,[6,14] $O_{II}$,[15] and $O_{III}$[16] are suggested to be the eligible candidates. A structural phase transition from α-$V_2O_5$ to metastable γ´-$V_2O_5$,[17] however, is also proposed with the increase in temperature to explain metallicity above the trasition temperature, as γ´-$V_2O_5$ resembles conductive Wadsley phase of $V_4O_9$.[18]

In the present study, we report semiconductor to metal transition of (001) oriented $V_2O_5$ nanorods, grown using vapour transport process. A transition from semiconducting to metallic behavior was recorded in the temperature dependent I-V measurement. The reversible temperature dependent Raman spectra were analyzed for the allowed vibrational modes of semiconducting and metallic $V_2O_5$ phases to understand the origin of the transition. The change in electronic band structure after the reduction of $V_2O_5$, prompting the metallicity in the sample, is discussed using temperature dependent Ultraviolet-Visible (UV-Vis) absorption spectroscopic studies and thermal activation energy calculations.

## 2. Experimental details

$V_2O_5$ nanorods were synthesized by vapor transport process using bulk $VO_2$ powder (Sigma-Aldrich, 99%) as source, placed in a high pure (99.99%) alumina boat at the center of the quartz tube reaction chamber, and flowing 20 sccm of Ar (99.9%) as carrier gas. Samples were grown on $SiO_2$/Si (100) using Au thin film (2 nm) as catalyst. The substrate was kept 5 cm away from the source normal to the stream of Ar. The synthesis was carried out at 1173K for 2 h.

Morphological analysis of the pristine sample was studied using a field emission scanning electron microscope (FESEM, SUPRA 55 Zeiss). The crystallographic studies were performed with the help of glancing incidence x-ray diffractometer (GIXRD; Bruker D8) using a Cu $K_\alpha$ radiation source ($\lambda$=1.5406 Å) with a glancing angle ($\theta$) of 0.5º in the $\theta$-$2\theta$ mode. A micro-Raman spectrometer (inVia, Renishaw, UK), in the back scattering configuration, was used for $Ar^+$ laser (514.5 nm) excitation with a diffraction gratings of 1800 gr.mm$^{-1}$ for monochromatization and a thermoelectric cooled charged coupled device (CCD) detector to study the vibrational modes. Electrical measurements were carried out by two Au coated contact tips. The activation energy $E_a$ was calculated from the slope of ln(R) vs. (1/T) plot, using the formula $R = R_0 \exp(-E_a/K_B T)$ where $K_B$ is the Boltzmann constant, $R_0$ is the resistance at temperature T=0K. Absorption spectra were recorded using an UV-Vis absorption spectrometer (Avantes) in the range of 200 to 700 nm. The Tauc's plots of the as-grown material, were drown using $(\alpha h\nu)^{1/n}$ vs. $h\nu$ with $n = 2$ for indirect band gap semiconductor (absorption co-efficient, $\alpha$; Planck's constant, $h$; frequency, $\nu$). Temperature dependent spectroscopic studies and electrical measurements were performed in the Linkam (THMS600) stage.

## 3. Results and discussions

The typical field FESEM image (Figure 1(a)) of as-grown sample shows rod like structure with an average diameter of 200-300 nm. The inset in figure 1(a) shows high magnification image of a typical single nanorod.

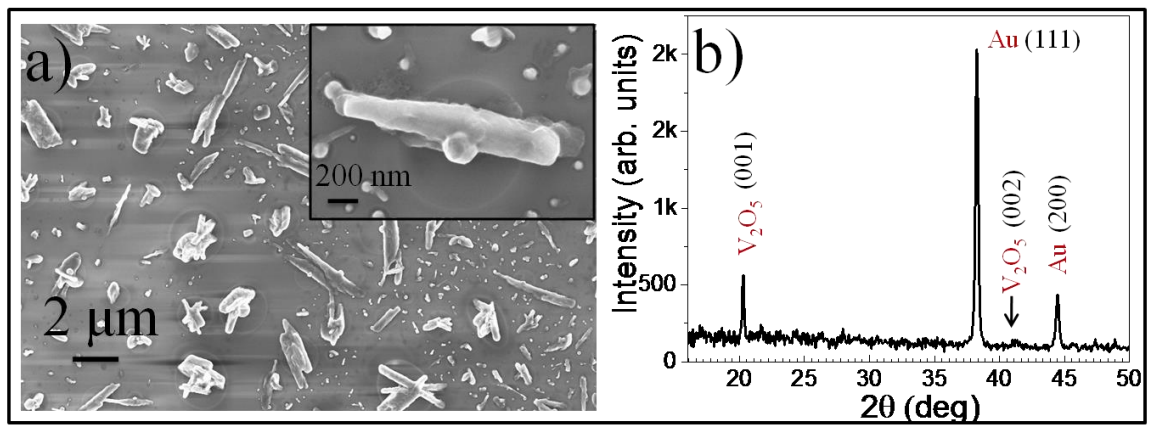

**Figure 1.** (a) FESEM image of as-grown nanorods. Inset shows a high resolution image of a typical nanorod (b) GIXRD spectrum of as-grown sample indicating crystalline planes corresponding to phases present.

The phase of the as-grown material was analyzed by GIXRD in the studies. Figure 1(b) shows the GIXRD data with peaks corresponding to the (*hkl*) planes of (001) and (002), confirming pure $V_2O_5$ phase (ICCD 00-041-1426) with unit cell dimensions $a$ = 11.51 Å, $b$ = 3.56 Å and $c$ = 4.37 Å grown with (001) crystalline orientation.[13] As we have used Au as catalyst for the nanostructure growth, the GIXRD spectrum shows peaks corresponding to (111) and (200) planes for Au (ICCD 00-004-0784) also.

Group theoretical analysis predicts twenty one Raman active modes for $V_2O_5$ at $\Gamma$ point, $7A_g+7B_{2g}+3B_{1g}+4B_{3g}$.[19] However, we observed eleven Raman modes for as-grown nanorods (Figure 2), which match with the reported data for $V_2O_5$.[19,20]

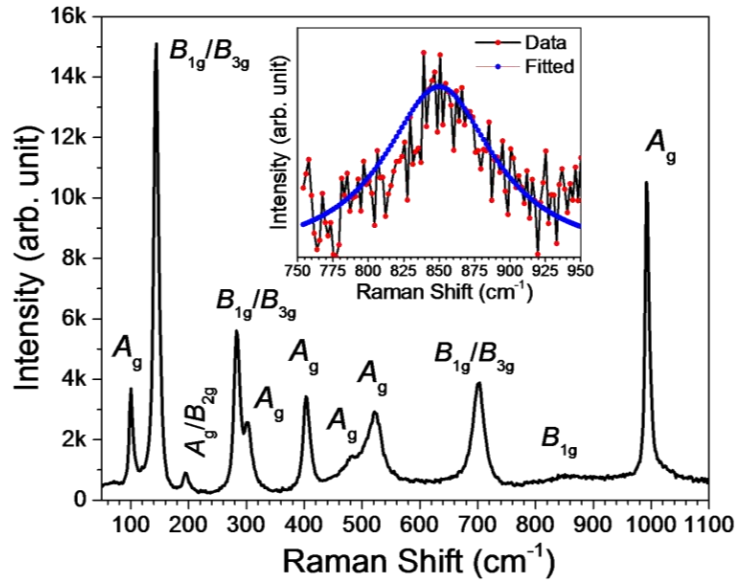

**Figure 2**. Raman spectrum of the as grown $V_2O_5$ nanorods with excitation of 514.5 nm at room temperature. The inset shows fitted peak at 850 cm$^{-1}$ corresponding to $B_{1g}$ mode.

The Raman peaks at 102 ($A_g$), 144 (either $B_{1g}$ or $B_{3g}$; $B_{1g}/B_{3g}$), 195 ($A_g/B_{2g}$), 283 ($B_{1g}/B_{3g}$), 301 ($A_g$), 403 ($A_g$), 483 ($A_g$), 523 ($A_g$), 701 ($B_{1g}/B_{3g}$), 850 ($B_{1g}$) and 993 ($A_g$) cm$^{-1}$ confirm the presence of pure $V_2O_5$ phase.[19] Orthorhombic $V_2O_5$ is made of distorted $VO_5$ pyramids sharing edges and corners having space group $P_{mmn}$ ($D_{2h}^{13}$).[21] The eleven observed Raman peaks can be assigned as follows. The highest frequency peak at 993 cm$^{-1}$ appears due to the stretching vibrational mode of V-$O_I$ bond along $Z$ direction (detail structure may be referred to supplementary information Figure S1). The peak at 850 cm$^{-1}$ (inset of Figure 2), observed experimentally for the first time, is predicted to originate because of antiphase stretching mode of V-O$_{II}$ bonds corresponding to the displacement of O$_{II}$ atoms along *X* direction.[19] The very weak intensity of the mode is due to pseudo-centrosymmetric nature of V-O$_{II}$-V bond.[20] Displacement of O$_{III}$ atoms in *Y* and *X* directions generates Raman modes at 701 cm$^{-1}$ (V-O$_{III}$-V antiphase stretching mode) and 523 cm$^{-1}$ ($d_4$ stretching vibration), respectively. The V-O$_{II}$-V bending deformation along *Z* direction (*c* axis) gives rise to Raman mode at 483 cm$^{-1}$. Modes at 403 and 283 cm$^{-1}$ can be attributed to oscillation of O$_I$ atoms along *X* and *Y* axes, respectively. One the other hand, displacement of O$_{II}$ atoms along *Z* axis gives rise to Raman peak at 301 cm$^{-1}$. The low frequency modes at 195, 144 and 102 cm$^{-1}$ correspond to the *X*, *Y* and *Z* displacements of the whole chain involving shear and rotations of the ladder like V-O$_{III}$ bonds.[20,22] The high intensity of 144 cm$^{-1}$ peak indicates the long range order of V-O layers in the *XY* plane.[20]

V$_2$O$_5$ is an indirect band gap semiconductor, with a gap value of about 2 eV.[23,24] Presence of two localized narrow split-off bands with a gap of ~ 0.7 eV at the bottom of conduction band is considered as the most interesting feature of its electronic band structure.[24] The upper split-off band is separated from the main conduction band by an additional gap of 0.35 to 0.5 eV.[23-25] However, V$_2$O$_5$ shows metallic behavior above the transition temperature.

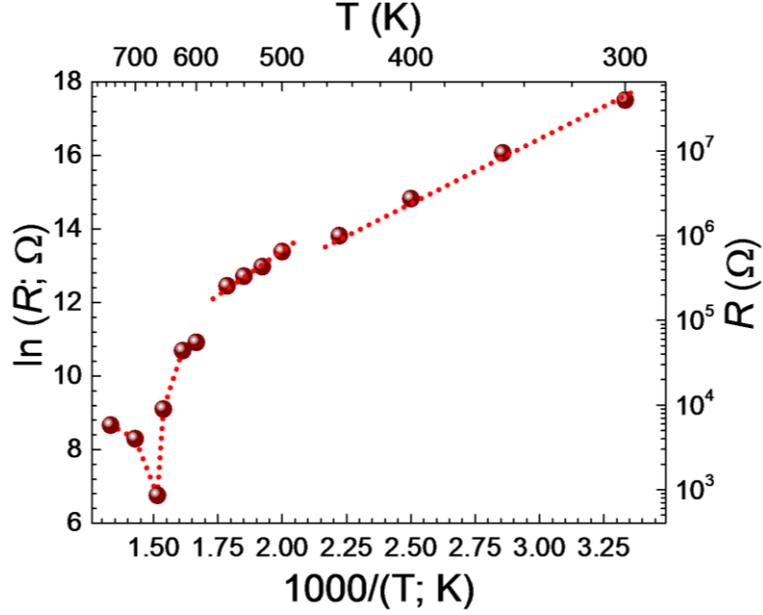

**Figure 3**. Change in the resistance with temperature in an ensemble of $V_2O_5$ nanorods

The semi log plot of resistance ($R$) vs. $T^{-1}$ shows (Figure 3) that resistance decreases exponentially up to 450K with increasing temperature, indicating a semiconducting behavior leading to an activation energy of 0.29 ± 0.01 eV, which is slightly higher than the previously reported value (0.26 eV) for single crystalline $V_2O_5$, measured in the temperature range of 200-270K.[26] The most possible reason for this low value of activation energy is due to the transition of electrons from localized split-off bands to main conduction band. In the temperature range of 500-550K, the slope of the plot is increased with an activation energy of 0.4 ± 0.01 eV, which indicates the downward shift of split-off bands with increasing temperature. Above 550K, it starts decreasing rapidly disobeying the semiconducting nature, and the plot shows metallic behavior of increasing resistance with increase in the temperature above 650K. As discussed earlier, the cause of the metallic behavior is still under debate.[5,6,15-17] The proposed metallic phase of metastable γ´-$V_2O_5$ in the structural phase transition model,[17] is reported to convert to a stable semiconducting α-$V_2O_5$ phase above 613K.[27] In the present study, however, the temperature dependent electrical measurement shows (Figure 3) that the metallic character of the grown nanorods sustains above 650K, which contradicts the formation of γ´-V$_2$O$_5$. Hence the reduction of V$_2$O$_5$ though O vacancy with no structural change is likely in producing the metallic phase above the transition temperature.[5] However, the role of specific O (supplementary information Figure S1) is still not clear in producing the metallic phase.

A temperature dependent Raman spectroscopic study (Figure 4(a)) was conducted to address the role of specific O in producing the metallic phase.

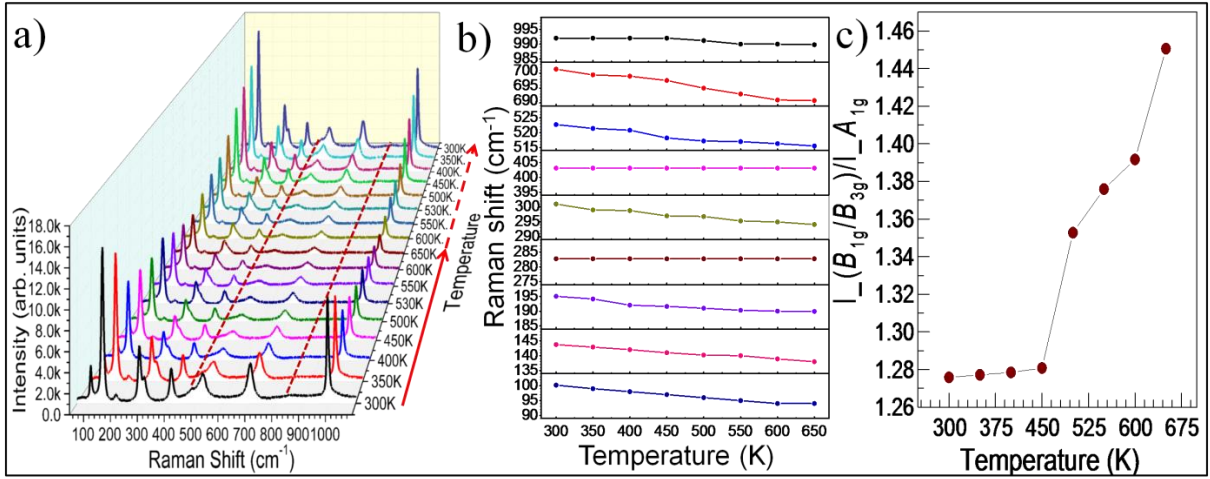

**Figure 4**. (a) Raman spectra of V$_2$O$_5$ nanorods with increasing (denoted by solid arrow) and decreasing (denoted by dashed arrow) temperature in the range of 300 to 650K (b) Change in vibrational frequency with temperature for each Raman mode and (c) Temperature dependence of the intensity ratio of the highest intense peak at 144 cm$^{-1}$ (I_($B_{1g}/B_{3g}$)) with the peak at 993 cm$^{-1}$ (I_$A_{1g}$).

In temperature dependent Raman study, the modes at 483 and 850 cm$^{-1}$ are observed to disappear completely above the transition temperature of 530K and reappear again at the same temperature while cooling down. Moreover, a softening for all Raman modes was recorded with the increase in temperature except for the mode frequencies at 283, 403 and 993 cm$^{-1}$. Temperature dependence (increasing) of frequencies for all the observed Raman modes is shown in figure 4(b). The peak positions of the Raman modes, except for the three above mentioned modes, are red shifted by an amount of 7 to 13 cm$^{-1}$ in the temperature range from 300 to 650K. The intensity of the peak centered at 993 cm$^{-1}$ is observed to decrease rapidly with the increase in temperature and regain almost its original intensity after cooling back to room temperature. The intensity ratio of the highest intense 144 cm$^{-1}$ ($B_{1g}/B_{3g}$) Raman mode with 993 cm$^{-1}$ ($A_{1g}$) mode with increasing temperature is shown in figure 4(c). The intensity ratio (I_($B_{1g}/B_{3g}$)/I_$A_{1g}$) increases rapidly above the temperature 450K indicating a significant decrease in the intensity of $A_{1g}$ mode centered at 993 cm$^{-1}$ as compared to that of $B_{1g}/B_{3g}$ mode of 144 cm$^{-1}$ around the transition temperature of 530K. The drastic fall in intensity of the $A_{1g}$ mode, which is responsible for the vibration of vanadyl $O_I$ atoms along *c*-axis alone, signifies the possible loss of $O_I$ atoms from the structure around the transition temperature. It may also be noted that the $B_{1g}/B_{3g}$ mode frequency at 144 cm$^{-1}$ is independent of the motion of $O_I$, so our inference about the role of $O_I$ influencing the intensity of $A_{1g}$ mode at 993 cm$^{-1}$ is mutually exclusive. Moreover, it is reported that the vanadyl $O_I$ is more prone to reduction, as the vacancy formation energy is lower for $O_I$ atoms than the other two coordinated oxygen atoms ($O_{II}$ and $O_{III}$) in $V_2O_5$ (001) oriented surface.[28,29] The modes at 483 and 850 cm$^{-1}$, arising due to V-$O_{II}$-V bending and stretching vibrations, respectively, disappeared above the transition temperature. It may be due to the relaxation of $V_2O_5$ structure initiated with the formation of vanadyl $O_I$ vacancy (schematic in the supplementary information Figure S2). If one of the vanadyl oxygen atoms are removed from the surface, the V atom in its near vicinity projects inward for relaxation and the next right vanadyl oxygen relax upward to make a stiffer interlayer bond with increased bond length (1.78 Å). Moreover, the V-$O_{II}$-V bond angle is also reported to increase to 178º leading to almost a linear bond.[28] So, the disappearance of the modes at 483 and 850 cm$^{-1}$ is in quite good agreement with the relaxation conditions. The absence of phonon softening with temperature for the Raman modes at 283, 403 and 993 cm$^{-1}$, which originates due to the *Y*, *X* and *Z* vibration of O$_I$ atoms, respectively, may also be attributed to the structural relaxation as the V-O$_I$ bonds become stiffer between the layers after the relaxation. The reversibility of the Raman modes with temperature can be explained by the excellent catalytic behavior of V$_2$O$_5$. Structural phase transition to γ´-V$_2$O$_5$ is further ruled out with the absence of Raman mode at 602 cm$^{-1}$,[30] in the studied temperature range of 300-650K.

To understand the origin of metallicity in V$_2$O$_5$, temperature dependent UV-Vis absorption of the as-grown V$_2$O$_5$ nanorods was studied. V$_2$O$_5$ is a semiconductor with an indirect band gap of ~2 eV corresponding to a transition from *R* to *Γ* point in the first Brillouin zone.[23,24] Two split-off bands with narrow bandwidth (0.45 to 0.75 eV) below the conduction band at *Γ* point are also reported.[24,25] Vanadyl O$_I$ plays an important role in creating these split-off conduction bands.[31,32] The Tauc's plot of the as-grown material, using *(αhv)$^{1/2}$* vs. *hv* for indirect band gap semiconductor (absorption co-efficient, *α*; Planck's constant, *h*; frequency, *v*), at different temperatures ranging from 300 to 650K are shown in figure 5.

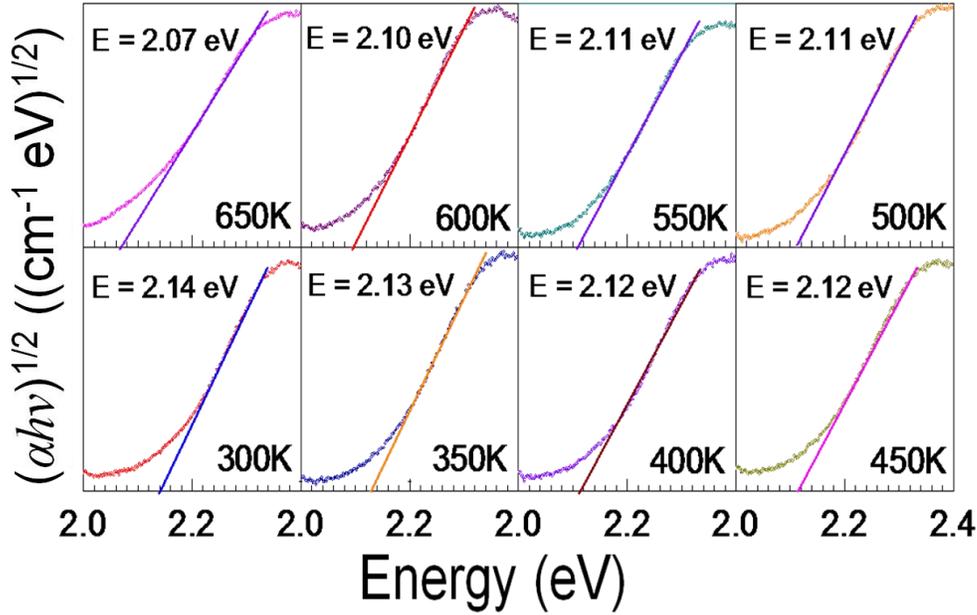

**Figure 5**. Tauc's plot of indirect band gap $V_2O_5$ nanorods using UV-Visible spectra at different temperatures. The slopes are drawn to determine the band gap value, as inscribed in the insets for the plots at different temperatures.

An indirect band gap of 2.14 eV, which matches with the previously reported value of 2.1 eV,[33] was recorded at 300 K. The band gap decreases with increase in temperature and is measured to attain a value of 2.07 eV at 650K. The decrease in band gap by an amount of 70 meV indicates downward shift of the split-off bands with increase in temperature, as also inferred from the thermal activation energy analysis (Figure 3). Inspired by the vanadyl $O_I$ vacancy, the split-off conduction bands are reported to approach deeper down from the conduction band at $\Gamma$ point due to the breakdown of $pd\pi$ bond between $O_I$ and nearest V atom.[14,31,32] Reduction of an oxygen atom donates two electrons back to the structure. The breakdown of $pd\pi$ bond between $O_I$ and nearest V atom results in delocalization of electrons towards neighboring V cation by occupying the partially filled V $3d$ states in the conduction band,[34] leading to the increase in the number of carriers in the conduction band. The increased number of carriers in the conduction band may be responsible for the decrease in resistance above the transition temperature and the observed metallic behavior. Thus, the decrease in band gap with increasing temperature supports our assumption of vanadyl $O_I$ vacancy from the $V_2O_5$ structure above the transition temperature.

## 3. Conclusion

In conclusion, orthorhombic $V_2O_5$ (001) nanorods were synthesized by vapor transport method. Temperature dependent electrical properties showed a transition of semiconducting to metallic behavior at temperature of 550K. Phonon softening behavior in the temperature dependent Raman spectra indicated loss of vanadyl oxygen ($O_I$) as the most possible reason for the observation of the metallic $V_2O_5$ phase above the transition temperature. Reduction of the band gap with increasing temperature, as observed in the Tauc's plot and thermal activation energy calculations, implies the downward movement of split-off bands from the conduction band. The downward shift of split-off bands with increase in temperature is due to breakdown of $pd\pi$ bond between $O_I$ and nearest V atom, inspired by the formation of vanadyl $O_I$ vacancy. The breakdown of $pd\pi$ bonds helps in accumulating electrons towards neighboring V atom to occupy the partially filled V *3d* bands, leading to the increase in the number of carriers in conduction band, which is finally made responsible for the decrease in the resistance and the observed metallic behavior, for the first time.

**AUTHOR INFORMATION:**

*Email Address: raktimabasu14@gmail.com; dhara@igcar.gov.in

**ACKNOWLEDGMENT**

One of us (RB) thanks DAE, India for financial support. We also acknowledge D. N. Sunitha, S. R. Polaki and A. Pasta of SND, IGCAR for their help in GIXRD, FESEM and electrical studies, respectively.

*Supplementary information*

**Role of Vanadyl Oxygen in Understanding Metallic Behavior of $V_2O_5$ (001) Nanorods**

*Raktima Basu,[*] Arun K. Prasad, Sandip Dhara,[*] and A. Das*

Nanomaterials and Sensors Section, Surface and Nanoscience Division, Indira Gandhi Centre for Atomic Research, Homi Bhabha National Institute, Kalpakkam−603 102, India

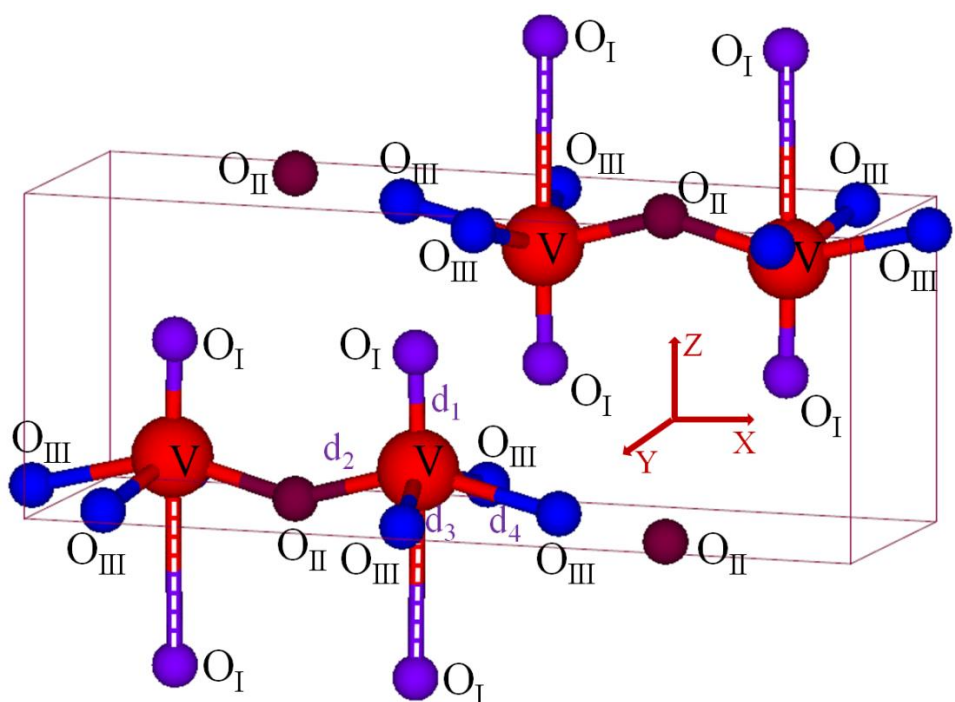

**Figure S1.** Schematic diagram of $V_2O_5$ unit cell

Figure S1 shows the schematic diagram of $V_2O_5$ unit cell. Orthorhombic $V_2O_5$ crystallizes in a layered structure perpendicular to the *Z*-axis consisting of $VO_5$ pyramids sharing their vertices and corners. There are three non-equivalent oxygen atoms in each unit cell (denoted as $O_I$, $O_{II}$, and $O_{III}$). $O_I$ is the terminal (vanadyl) oxygen with two different bond lengths. One of them is strong and short V-$O_I$ bond with length 1.577 Å ($d_1$). Another one is large and weak Van der Waals type, which connects the two adjacent layers in the $V_2O_5$ structure, with a bond length of 2.793 Å (showed by dotted white lines). Both of these vanadyl oxygen atoms orient almost along the *c*-axis. The two fold coordinated bridging oxygen ($O_{II}$) connects two adjacent V atoms with V-O$_{II}$ bond length of 1.78 Å ($d_2$). The ladder shaped O$_{III}$ atoms are the three-fold coordinated oxygen with three different V-O$_{III}$ bond lengths of 1.88 ($d_3$), 1.88 ($d_3$), and 2.02 Å ($d_4$) [Ref: 13 in the main text].

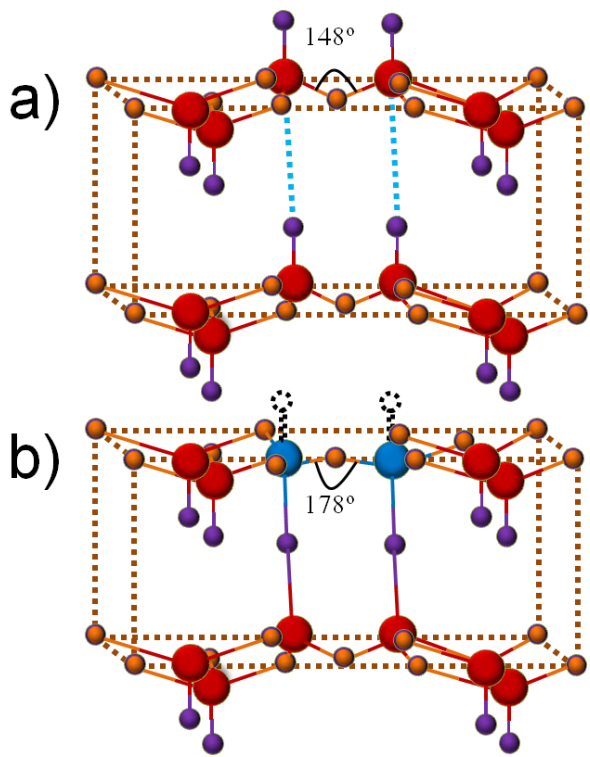

**Figure S2.** Schematic diagram of V$_2$O$_5$ unit cell (a) below and (b) above transition temperature

Figure S2 shows the structure of V$_2$O$_5$ unit cell below (Figure S2 (a)) and above (Figure S2 (b)) the transition temperature. If one of the vanadyl oxygen atoms are removed from the surface, the V atom in its near vicinity projects inward for relaxation and the next right vanadyl oxygen relax upward to make a stiffer interlayer bond with increased bond length (1.78 Å). Moreover, the V-O$_{II}$-V bond angle is also reported to increase to 178º leading to almost a linear bond. [Ref: 28 in the main text].